\documentclass[lettersize,journal]{IEEEtran}
\usepackage{amsmath,amsfonts}
\usepackage{algorithmic}
\usepackage{algorithm}
\usepackage{array}
\usepackage[caption=false,font=normalsize,labelfont=sf,textfont=sf]{subfig}
\usepackage{textcomp}
\usepackage{stfloats}
\usepackage{url}
\usepackage{verbatim}
\usepackage{graphicx}
\usepackage{cite}
\usepackage{hyperref}
\usepackage{cleveref}
\crefname{equation}{Eq.}{Eq.} 
\crefrangelabelformat{equation}{(#3#1#4)--(#5#2#6)}
\crefname{figure}{Fig.}{Fig.}
\crefname{table}{Table}{Table}
\crefname{section}{Section}{Section}
\newcommand{\flag}[1]{#1}
\begin{document}

\title{Experimental End-to-End Optimization of Directly Modulated Laser-based IM/DD Transmission}

\author{Sergio Hernandez, Christophe Peucheret,~\IEEEmembership{Member,~IEEE}, Francesco Da Ros,~\IEEEmembership{Senior Member,~IEEE}, Darko Zibar
\thanks{This work was supported by VILLUM FONDEN under Grants VI-POPCOM VIL54486, and OPTIC-AI VIL29334. \textit{(Corresponding author: Sergio Hernandez.)}\\
Sergio Hernandez, Francesco Da Ros and Darko Zibar are with the Department of Electrical and Photonics Engineering, Technical University of Denmark, 2800 Lyngby, Denmark. (e-mail: shefe@dtu.dk)\\
Christophe Peucheret is with Univ. Rennes, CNRS, FOTON - UMR6082, 22305 Lannion, France}
\thanks{Manuscript received April 19, 2021; revised August 16, 2021.}}

\markboth{Journal of \LaTeX\ }%
{Shell \MakeLowercase{\textit{et al.}}: A Sample Article Using IEEEtran.cls for IEEE Journals}


\maketitle

\begin{abstract}
Directly modulated lasers (DMLs) are an attractive technology for short-reach intensity modulation and direct detection communication systems. However, their complex nonlinear dynamics make the modeling and optimization of DML-based systems challenging. In this paper, we study the end-to-end optimization of DML-based systems based on a data-driven surrogate model trained on experimental data. The end-to-end optimization includes the pulse shaping and equalizer filters, the bias current and the modulation radio-frequency (RF) power applied to the laser. The performance of the end-to-end optimization scheme is tested on the experimental setup and compared to 4 different benchmark schemes based on linear and nonlinear receiver-side equalization. The results show that the proposed end-to-end scheme is able to deliver better performance throughout the studied symbol rates and transmission distances while employing lower modulation RF power, fewer filter taps and utilizing a smaller signal bandwidth.
\end{abstract}


\section{Introduction}

The explosive growth of internet services and data traffic, driven by the surge in cloud computing and artificial intelligence (AI) applications, is placing considerable pressure on the global communications infrastructure \cite{10273773}. Specifically, short-reach intra-data center interconnects must continue to increase data throughput while maintaining low costs. Intensity modulation and direct detection (IM/DD) is the dominating technology for such interconnects \cite{10238466, app13148125, 10526441}. Directly modulated lasers (DMLs) are an attractive technology for short-reach IM/DD optical links, offering low energy consumption, compact size, and affordability \cite{10528667, 10438718, 10080935, 10323176, 10041217, 10261204}. This makes them well-suited for communication systems demanding high data rates at minimal cost. DMLs are simpler and potentially more efficient than alternative transmitter technologies \cite{10041217}. However, the limited modulation bandwidth of DMLs can restrict their achievable symbol rates \cite{Sudo:21, 9874980, 8570831}. The interaction between modulation-induced frequency chirp and chromatic dispersion (CD) is also among the main limiting factors in the implementation of DML-based IM/DD systems \cite{10238466}. As throughput requirements increase beyond 100~Gbps per lane \cite{10238466}, these limitations become more challenging. Operating DMLs at symbol rates beyond 50~GBd, necessary for high data throughput, can introduce impairments such as nonlinear waveform distortion and more severe frequency chirping \cite{575810, 10925524, Zhang:25}. These impairments can reduce the effective signal-to-noise ratio (SNR) of the signal at the receiving end, limiting transmission distance and achievable data rates.

Traditionally, digital signal processing (DSP) techniques for mitigating impairments in DML-based systems have focused on receiver-side equalization (EQ) or a combination of digital pre-distortion (DPD) at the transmitter and EQ at the receiver \cite{Kottke:17, Xu:20, 10238466, 8570831, 8259239, 10528667}. These approaches optimize only one end of the system at a time and thus they may lead to suboptimal solutions compared to joint optimization techniques \cite{10839011}. End-to-end (E2E) learning \cite{8054694, Karanov:20, 10824218, Hernandez:24} offers a promising approach to jointly optimizing the transmitter and receiver, potentially leading to optimal DSP configurations in terms of signal quality and power consumption. E2E learning in communication systems involves representing the transmitter and receiver as a single autoencoder (AE) artificial neural network (ANN) built around the transmission channel \cite{10093964, 10124361}. This approach allows for the simultaneous optimization of the DSP at the transmitter (constellation shaping, pulse shaping, DPD) and receiver (EQ, symbol detection) while keeping the parameters of the channel model fixed.



Applying E2E learning to DML-based systems is hindered by the physical model of the laser \cite{Agrell_2024, 10093964, 10530894}. The DML large-signal dynamics are governed by nonlinear laser rate equations \cite{7972948}, which describe the interplay between carrier density and photon density in the cavity. These equations lack a closed-form large-signal solution, requiring the use of iterative numerical solvers to simulate the DML's output. This simulation step is not directly differentiable, which prevents the backpropagation of gradients from the receiver to the transmitter and stalls the training process \cite{10059125, 10261204}. Data-driven surrogate models based on machine learning offer a solution to the differentiability challenge posed by DMLs \cite{Hernandez:24}. These models learn a differentiable representation of the DML dynamics based on input/output sequences. Data-driven models create a viable, differentiable pathway for backpropagation where analytical models fall short, unlocking the potential of end-to-end, gradient-based optimization for DML-based systems. Data-driven models can also be faster substitutes for computationally expensive ordinary differential equation (ODE) solvers \cite{10382548} and derivative-free gradient approximators in parameter optimization \cite{9513589}. Based on the gradients provided by the differentiable surrogate model, E2E learning allows the joint optimization of an arbitrarily large space of parameters that affect the performance of the system.



In this paper, we demonstrate the surrogate-based E2E optimization of transmitter (TX) and receiver (RX) DSP and laser driving configurations over different lengths of standard single mode fiber (SSMF). The proposed data-driven surrogate model is trained to capture the chirp-CD interaction along the different fiber lengths, allowing the E2E optimization to mitigate its impact on performance. This is done by training a separate model for each transmission distance and baudrate. Our approach allows the E2E system to optimize the DSP (constellation shaping, pulse shaping, EQ and symbol detection) and laser-driving configurations (bias and modulation RF power) for every scenario. Thus, we extend our previous work in \cite{Hernandez:25} \flag{in three different ways. Firstly, while \cite{Hernandez:25} only addressed a back-to-back system, here we also study the performance of the system under transmission through 1 and 2 km of standard single-mode fiber (SSMF). This additional analysis assesses the system's capability to mitigate undesirable effects from the interaction between chromatic dispersion and frequency chirp. Second, we bypass the digital bandwidth limitation present in our previous work, simplifying the structure of the surrogate model. Finally, we analyze and interpret the TX and RX filters obtained through numerical optimization. The analysis clarifies the trade-offs taken by the E2E scheme and explains the results achieved.}
\flag{Although the achieved throughput does not reach that of state-of-the-art DMLs due to device limitations, the presented results aim to validate the proposed methods in terms of the potential performance improvements.} 
The proposed E2E system is benchmarked experimentally against two more traditional RX-side equalization based on root-raised cosine (RRC) pulse-shaped signals. The two RRC-shaped benchmark equalization strategies are based on (1) a linear (finite impulse response, FIR) and (2) a nonlinear (2$^\mathrm{nd}$ order Volterra series, VNLE) equalizer to test the E2E system performance against both linear and nonlinear receiver-side equalization. We demonstrate, for the first time to our knowledge, the joint optimization of DSP and laser driving configurations in an experimental transmission setting. The E2E system achieves better error performance and eye opening than RX equalization across the studied fiber lengths. This improvement is realized while utilizing less bandwidth and employing lower modulation RF power and fewer filter taps than the RX-side equalization benchmarks.

\begin{figure*}[ht]
    \centering
    \includegraphics[width=\linewidth]{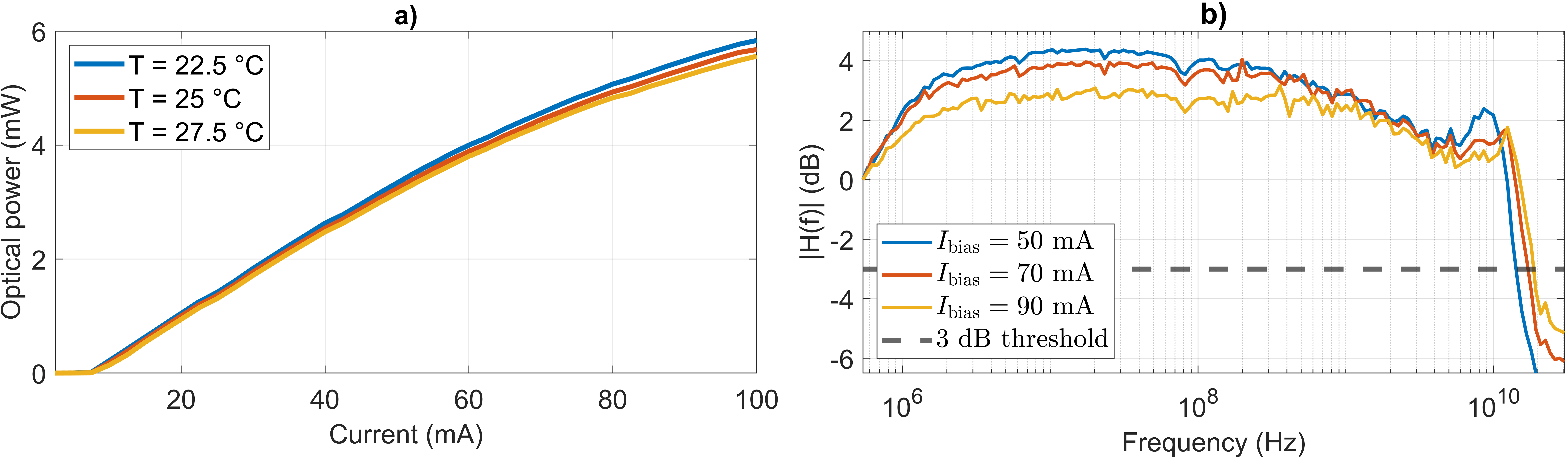}
    \caption{a) Measured light-current (L-I) curve and b) small-signal modulation response $H(f)$ of the utilized NEL NLK1551SSC DML.}
    \label{fig:l_i_h21}
\end{figure*}

The rest of the paper is structured as follows. \cref{sec:dml} describes the characteristics of the utilized DML. \cref{sec:methods} explains the methodology to build the surrogate model of the laser. \cref{sec:link} describes how the system parameters are optimized once the surrogate model is trained. \cref{sec:results} shows the experimental results of the E2E and benchmark setups in terms of symbol error rate (SER), eye diagrams and RX-side signal spectra. The results are summarized in \cref{sec:conc}. 

\begin{figure}[hb]
    \centering
    \includegraphics[width=\linewidth]{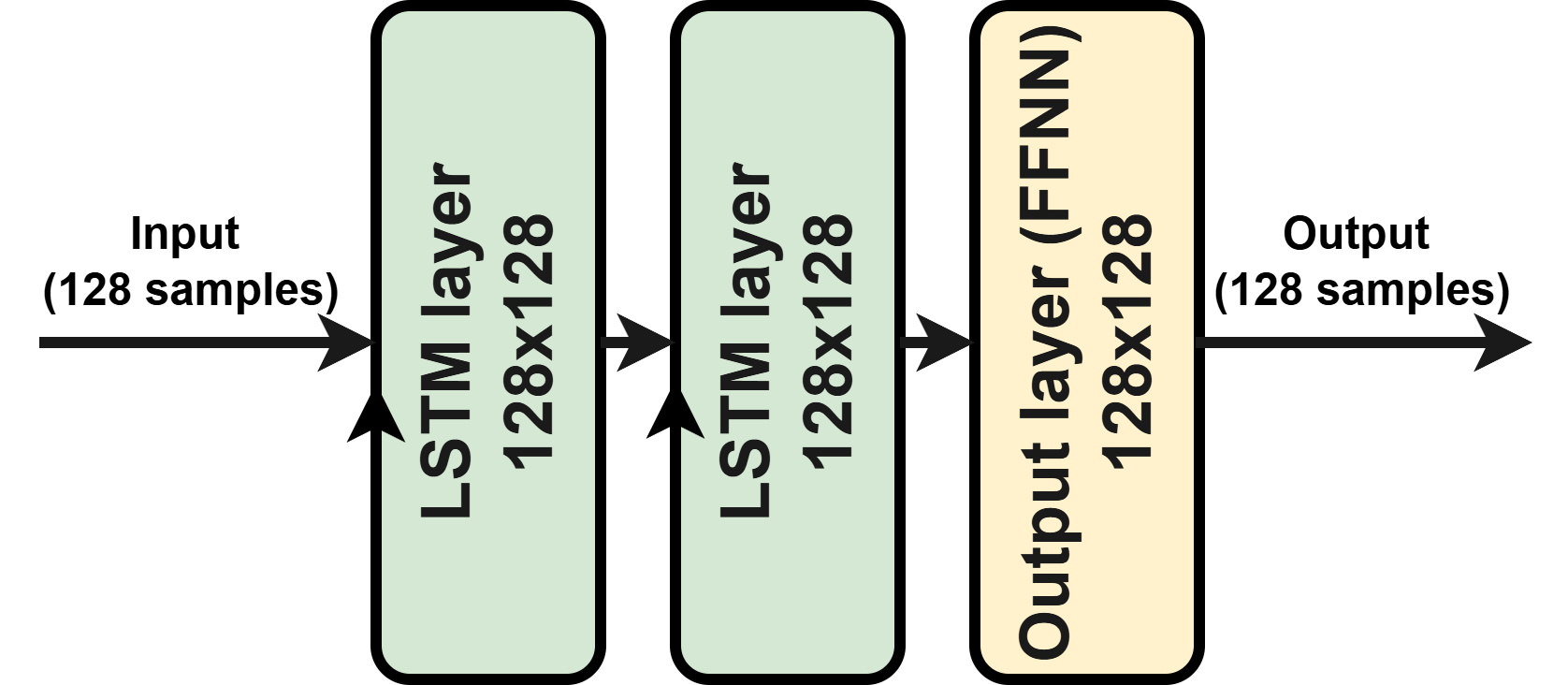}
    \caption{Structure of the surrogate model. The two numbers in each layer represent the input and output size of the layer, respectively. The vertical arrows represent the recurrence of LSTM layers. Acronyms: \textbf{LSTM}: Long-Short Term Memory; \textbf{FFNN}: feedforward neural network.}
    \label{fig:network}
\end{figure}

\section{DML characteristics} \label{sec:dml}
The characteristics of the laser play an important role in the error performance of DML-based IM/DD links, primarily due to the laser's strong nonlinear dynamics, limited bandwidth, and significant frequency chirping. The utilized NEL NLK1551SSC DML is designed for 10 Gbps links for up to 50~km, featuring a typical nominal external efficiency of 0.15 W/A. \flag{The bias current, or direct current component of the signal to the laser $I_\mathrm{bias}$ is limited to a maximum of 100~mA.} The measured L-I curves of the utilized laser at three different controlled temperatures $T \in [22.5, 25, 27.5]$~°C are shown in \cref{fig:l_i_h21}a. The threshold current $I_\mathrm{th}$ of the laser is measured to be 7.5~mA. The three L-I curves show a similar pattern in their slope (differential external efficiency, $\eta_d$), where the value of $\eta_d$ is maximum around 52.5~mA (0.067~W/A) and decreases for both higher and lower $I_\mathrm{bias}$ values. The small-signal modulation responses measured for three different $I_\mathrm{bias} \in [50, 70, 90]$~mA and  $T = 25$~°C are plotted in \cref{fig:l_i_h21}b. The responses are measured by sweeping the frequency of an input sinusoidal electrical current and recording the corresponding magnitude of the modulated optical output power to characterize the bandwidth of the laser. The peak-to-peak current of the sinusoidal input signal is 8~mA. The measured 3-dB bandwidth for  $I_\mathrm{bias} = [50, 70, 90]$~mA are $[14.2, 17.1, 18.7]$~GHz, respectively.

This leads to the significant difference in the low frequency response of the 3 curves. \cref{fig:l_i_h21} also hints at a trade-off between extinction ratio and modulation bandwidth (increasing with higher $I_\mathrm{bias}$ in \cref{fig:l_i_h21}b), with the $I_\mathrm{bias} \approx 50$~mA region offering a higher extinction ratio while delivering reduced bandwidth compared to higher bias currents.

\section{Experimental modeling} \label{sec:methods}

\begin{figure*}[t]
    \centering
    \includegraphics[width=\linewidth]{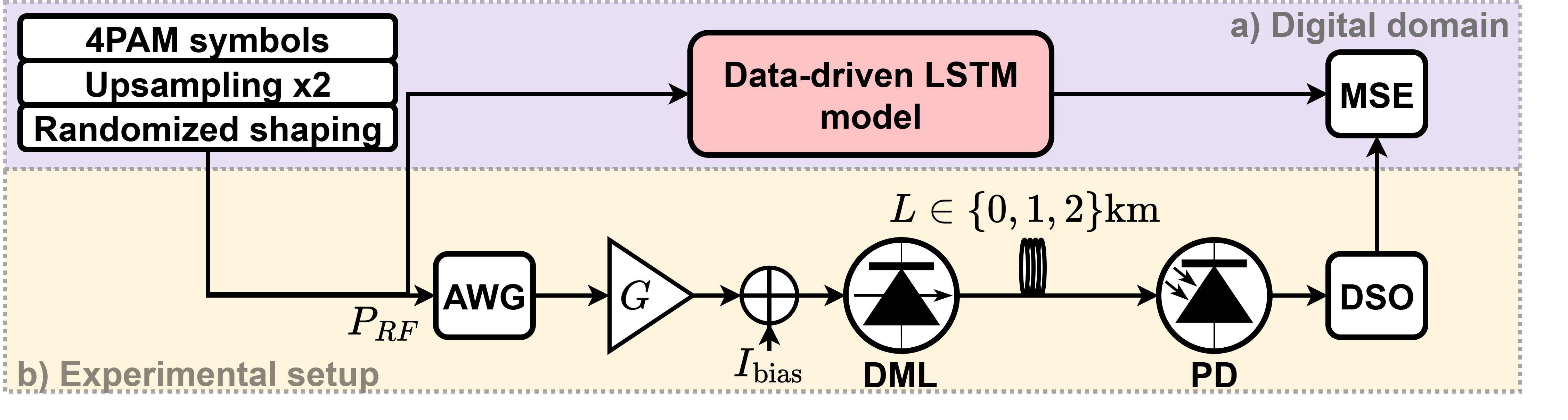}
    \caption{Block diagram of the proposed experimental modeling approach. The digital domain \flag{(a)} is designed to build a data-driven model resembling the dynamics of the experimental setup (b). Acronyms and symbols: \textbf{4PAM}: 4-Pulse Amplitude Modulation; \textbf{AWG}: arbitrary waveform generator; $\mathbf{P_\mathrm{RF}}$: modulation RF power; $\mathbf{G}$: RF amplifier gain; $\mathbf{I_\mathrm{bias}}$: bias current; \textbf{DML}: directly modulated laser; $\mathbf{L}$: standard single-mode fiber length; \textbf{PD}: photodetector; \textbf{DSO}: digital storage oscilloscope; \textbf{LSTM}: Long-Short Term Memory; \textbf{MSE}: mean squared error.}
    \label{fig:modelsetup}
\end{figure*}

To enable end-to-end optimization, we first train a differentiable data-driven model that captures the deterministic impairments in the link based on averaged experimental data. \flag{The model is intended as part of the offline calibration of the transmission system, and it does not need to run in real time. Therefore,} averaging is performed to discard the stochastic components (noise) in the signal, simplifying the modeling task. The model imitates the static and dynamic response of the link from TX to RX, including laser dynamics, fiber-induced distortions, and photodetector nonlinearities. Given the memory effects and nonlinearity introduced by the DML, the structure of the data-driven model must include memory mechanisms and nonlinear activations that allow it to infer the response of the link accurately. Based on our previous work in DML modeling \cite{10382548}, we choose a long short term memory (LSTM) network structure for the task. LSTMs are recurrent neural networks (RNNs) aimed at the inference of sequential data with long temporal dependencies, making them ideal candidates to model the memory effects and nonlinearity associated with the response of DMLs. 
The core principle of LSTMs is the use of hidden states $\mathbf{h}_t$, that are updated for every time sample $t$ to retain information about the past inputs and outputs in order to provide temporal context. The balance between past and present information stored in $\mathbf{h}_t$ is determined through matrix multiplication and nonlinear activation, allowing LSTMs to operate as universal approximators \cite{SIEGELMANN1995132}. Thus, the calculation of the output, or output gate $\mathbf{o}_t$ of an LSTM can be expressed as

\begin{equation}
    \mathbf{o}_t = \sigma\left(\mathbf{W}_o \begin{bmatrix} \mathbf{x}_t \\ \mathbf{h}_{t-1} \end{bmatrix} + \mathbf{b}_o \right),
\end{equation}

\noindent where $\sigma$ is the sigmoid function $\sigma(x) = \frac{1}{1 + e^{-x}}$, $\mathbf{W}_o$ are the learnable network weights, $\mathbf{x}_t$ is the time window of input samples, $\mathbf{h}_{t-1}$ is the hidden state matrix from the previous time step and $\mathbf{b}_o$ are the learnable biases of the network. The joint training of $\mathbf{W}_o$ and $\mathbf{b}_o$ determines the optimal balance between past and present data for modeling the system's dynamics. The structure of the network used as surrogate model is shown in \cref{fig:network}. The input and output size of every layer is set to 128 samples. The network is built using 2 stacked LSTM layers and a linear output layer. The use of stacked LSTM neurons allows each LSTM neuron to specialize in capturing different levels of dependencies, with some neurons focusing on short-term features and some other on longer-term ones. As in most NNs, the optimal network size and depth must be numerically searched while finding a compromise between inference accuracy and training/inference time. 

The experimental modeling strategy is shown in \cref{fig:modelsetup}a. The dataset employed for the model training is obtained by generating digital waveforms (before digital to analog conversion in the AWG) and propagating them through an experimental setup, shown in \cref{fig:modelsetup}b. The digital waveforms, used as input to the surrogate model, are generated based on equiprobable 4-pulse amplitude modulation (4PAM) symbols, shaped using a 2-tap FIR filter with randomized taps at 2 samples per symbol (SpS). The FIR filter length and the oversampling rate of 2 SpS were chosen to emulate the bandwidth limitations inherent in cost-constrained short-reach links. The FIR pulse shaping is re-randomized every 1024 symbols. $P_\mathrm{RF}  \in [-4, 2]$ dBm and $I_\mathrm{bias} \in [50, 100]$ mA are also randomized within their respective ranges, drawing their values from a uniform distribution. The modulation RF power $P_\mathrm{RF}$ applied to the DML is defined as \cref{eq:prf}.

\begin{equation}
P_\mathrm{RF} = \frac{1}{N_\mathrm{samps}} \sum_{n=1}^{N_\mathrm{samps}} |x_\mathrm{TX}[n]|^2,
\label{eq:prf}
\end{equation}

\noindent where $\mathbf{x}_\mathrm{TX}$ is the digital pulse-shaped signal to be transmitted, $N_\mathrm{samps}$ is the number of samples of $\mathbf{x}_\mathrm{TX}$. The model is trained on a dataset with varied waveforms, input powers ($P_\mathrm{RF}$), and bias currents ($I_\mathrm{bias}$) to ensure its robustness against fluctuations in these parameters. A 1024-sample pilot sequence is introduced before the 4PAM waveforms in order to allow simple symbol synchronization on the receiver side through autocorrelation. The experimental setup is organized into signal generation, transmission, and reception stages. The generation stage begins with a Keysight M8195A AWG (sampling rate: 65~GSa/s, bandwidth: 25~GHz, max $P_\mathrm{RF}$: 4~dBm). Its output is amplified by an SHF 806E RF amplifier ($G = 13$~dB, \flag{bandwidth: 38~GHz}) and drives a NEL NLK1551SSC DML, which is temperature-stabilized at 25\textdegree C and biased using a DC power supply generating $I_\mathrm{bias}$. For transmission, the optical signal propagates through one of three different lengths ($L$) of SSMF: back-to-back, 1~km, or 2~km. At the receiver, a Coherent Corp. XPDV2020R photodetector (PD) with a nominal bandwidth of 50~GHz \flag{and a nominal responsivity of 0.65 A/W} converts the signal. Finally, an Agilent DSOX93304Q oscilloscope (DSO sampling rate $R_\mathrm{sa}$: 80 GSa/s, bandwidth: 33~GHz) digitizes the received waveform experimentally. To match the transmitter hardware, the input sequences are linearly interpolated to 65~GSa/s before being sent to the AWG. However, the surrogate model is trained using linear interpolation to 80~GSa/s, matching the input and output sampling frequency to $R_\mathrm{sa}$.

\begin{figure*}[hb!]
    \centering
    \includegraphics[width=\linewidth]{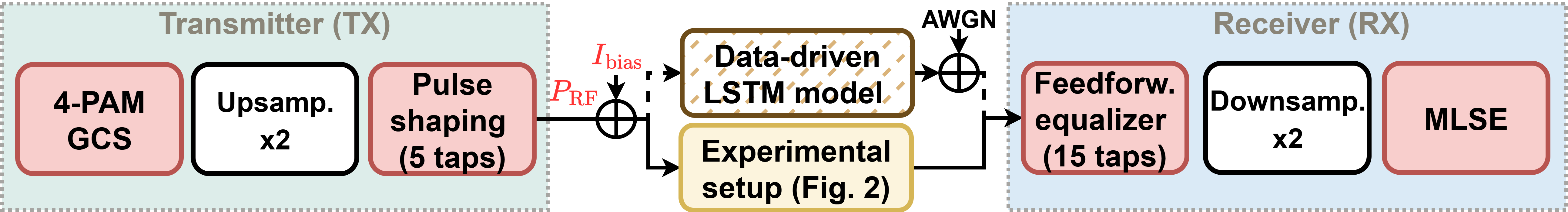}
    \caption{Block diagram of the E2E scheme parameter optimization. The elements optimized by the scheme are highlighted in red. Acronyms and symbols: \textbf{4PAM}: 4-Pulse Amplitude Modulation; $\mathbf{P_\mathrm{RF}}$: modulation RF power; $\mathbf{I_\mathrm{bias}}$: bias current; \textbf{LSTM}: Long-Short Term Memory; \textbf{MLSE}: maximum likelihood sequence estimation.}
    \label{fig:dsp_e2e}
\end{figure*}

\begin{table}[ht]
\centering
\caption{NRMSE loss of the trained DML surrogate models}
\begin{tabular}{|c|c|c|}
\hline
\multicolumn{1}{|c|}{\textbf{$L$ (km)}} & \multicolumn{1}{c|}{\textbf{20 GBd}} & \multicolumn{1}{c|}{\textbf{30 GBd}} \\ \hline
0                                     &      $1.07 \times 10^{-2}$                                &     $9.92 \times 10^{-3}$                                  \\ \hline
1                                     &         $8.84 \times 10^{-3}$                                 &     $8.73 \times 10^{-3}$                                     \\ \hline
2                                     &           $8.73 \times 10^{-3}$   &     $7.94 \times 10^{-3}$                                     \\ \hline
\end{tabular}
\label{tab:nrmse}
\end{table}

Once the waveforms have been propagated, the resulting distorted signal is captured 25 times and averaged in order to reduce stochastic effects, that cannot be captured by the LSTM model. The number of averages was chosen to balance sufficient noise suppression with a practical acquisition time. Once the full input/output dataset has been captured, averaged and synchronized, the data-driven LSTM model is trained using mean square error (MSE) between predicted and reference sequences as a loss function. The total size of the dataset is $10^7$ samples per baudrate and $L$ considered. \flag{Although training one model per baudrate and $L$ may increase operational complexity, it yields better modeling accuracy for each studied scenario.} The training is based on an Adam optimizer with learning rate $\mu_\mathrm{model} = 1 \times 10^{-4}$ and exponential decay rates $\beta_1 = 0.999, \beta_2 = 0.99$. The obtained surrogate model testing normalized mean square errors (NRMSEs) are listed in \cref{tab:nrmse}. The surrogate model is able to deliver normalized mean errors around the 1\% error mark throughout the $R_s$ and $L$ studied. However, the accuracy of the gradients provided by the surrogate model cannot be directly captured using MSE as a metric, and it must be assessed through the performance of the E2E optimization scheme.

\section{Link optimization} \label{sec:link}

\subsection{End-to-End scheme}
The architecture of the TX and RX DSP is shown in \cref{fig:dsp_e2e}. The blocks marked in red are optimized as part of the E2E learning scheme. The link optimization is performed offline based on the previously-trained surrogate model of the link, obtained from experimental data. The link optimization is based on 2 stages: the optimization and the validation phase. The optimization stage does not involve the experimental setup, as the surrogate model captures its deterministic dynamics. 
The E2E parameter optimization involves several blocks. Geometric constellation shaping (GCS) is used to optimize the constellation points for maximum information rate, pulse shaping tailors the transmitted pulse to control intersymbol interference, and equalization compensates for distortions from transmission. The E2E optimization also includes tuning the DML driving configuration, $I_\mathrm{bias}$ and $P_\mathrm{RF}$, to balance the extinction ratio against the modulation bandwidth. The TX DSP consists of a sequence of uniformly distributed 4PAM symbols, encoded as one-hot encoded vectors. The 4PAM symbols are shaped using learnable GCS, finding the optimal intensity level for each of the 4PAM symbols. Each 4PAM symbol is then upsampled to 2 SpS and pulse-shaped using a learnable 5-tap FIR filter. We determine the lengths of the pulse shaping and equalization FIR filters using hyperparameter optimization, subject to the constraint that their combined length cannot exceed 20 taps. The pulse-shaped discrete sequence is soft-clipped between 0 and 0.7~V peak-to-peak using the hyperbolic tangent (tanh) function. $I_\mathrm{bias}$ is treated as an optimizable parameter and clipped between 50 and 100~mA using a tanh function. The clipping ensures that the model does not converge to values that could damage the DML. The final signal to be transmitted $\mathbf{x}_\mathrm{TX}$ is linearly interpolated to 80~GSa/s and used as input to the surrogate model. The control of the signal amplitude and bias enables the optimization scheme to indirectly determine $P_{\mathbf{x}_\mathrm{TX}}$, allowing to find the optimal transmission $P_\mathrm{RF} = P_{\mathbf{x}_\mathrm{TX}}$.

The noise of the experimental setup is included in the model by adding additive white Gaussian noise (AWGN) equivalent to the estimated effective SNR ($\mathrm{SNR}_\mathrm{est}$) in the link. $\mathrm{SNR}_\mathrm{est}$ is calculated according to:

\begin{equation}
    \mathrm{SNR}_\mathrm{est} = \frac{P_{\mathbf{y}_\mathrm{avg}}}{P_n},
    \label{eq:snrest}
\end{equation}

\begin{figure}[ht!]
    \centering
    \includegraphics[width=\linewidth]{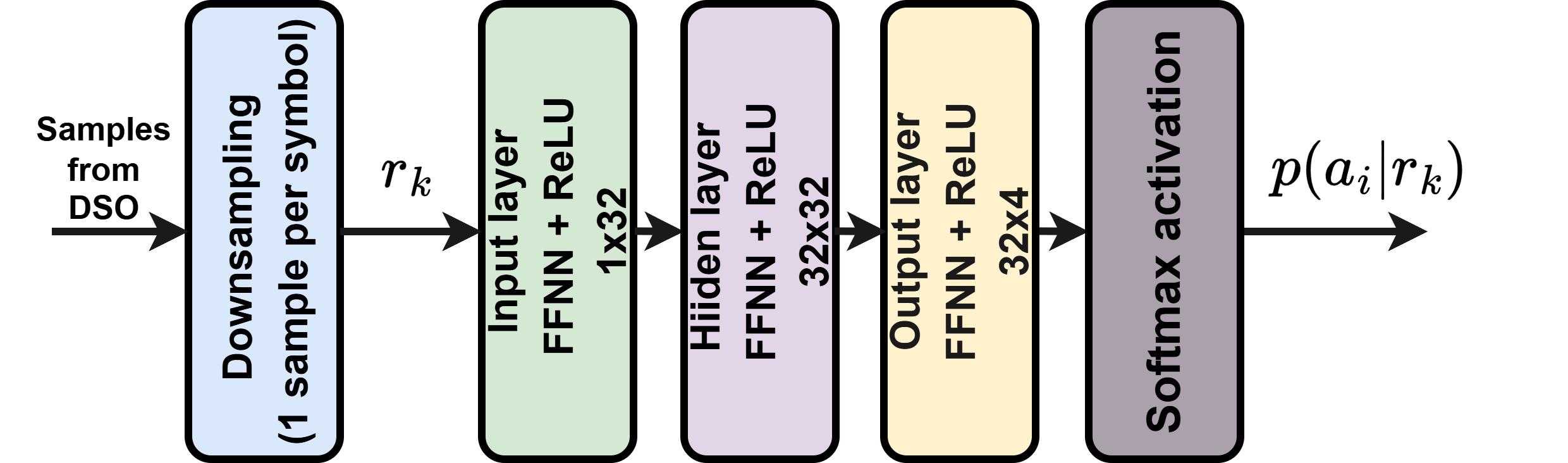}
    \caption{Structure of the maximum likelihood sequence estimation (MLSE) within the end-to-end (E2E) optimization. The two numbers in each layer represent the input and output size of the layer, respectively. Acronyms: \textbf{FFNN}: feedforward neural network; \textbf{ReLU}: rectified linear unit.}
    \label{fig:mlse}
\end{figure}

\noindent where $\mathbf{n} \approx \mathbf{y}_\mathrm{avg} - \mathbf{y}_\mathrm{raw}$ is the estimated noise sequence, $\mathbf{y}_\mathrm{avg}$ is the output training sequence after averaging and $\mathbf{y}_\mathrm{raw}$ is one of the 25 sequences used during the averaging of $\mathbf{y}_\mathrm{avg}$. $P_{\mathbf{y}_\mathrm{avg}}$ and $P_n$ represent the average power of $\mathbf{y}_\mathrm{avg}$ and $\mathbf{n}$, respectively. Once $\mathrm{SNR}_\mathrm{est}$ is calculated, the variance of the AWGN noise is calculated as: $\sigma^2_\mathrm{AWGN} = P_{\mathbf{y}_\mathrm{mod}} / \mathrm{SNR}_\mathrm{est}$, where $\mathbf{y}_\mathrm{mod}$ is the output sequence of the surrogate model. The noisy sequences predicted by the surrogate model are then processed by the RX DSP. The received 80~GSa/s sequences are linearly downsampled to 2 SpS before equalization. The receiver-side equalization is based on a trainable 15 $T/2$-spaced-tap FFE. The equalized signal is then downsampled to 2 SpS and detected using 1-tap maximum likelihood sequence estimation (MLSE) based on the trainable feedforward neural network (FFNN) \flag{shown in \cref{fig:mlse}}. The MLSE FFNN is based on a $1\times 32$ ReLU-activated feedforward input layer, followed by a hidden $32 \times 32$ feedforward layer with identical activation. The output $32 \times 4$ feedforward layer is activated by a softmax function. The MLSE FFNN serves as an approximation to the non-continuous, non differentiable thresholding function, delivering symbol probabilities as an output (soft decision). The link optimization uses categorical cross-entropy $H$ as loss function:

\begin{equation}
    H(\mathbf{a}, \mathbf{r}) = - \frac{1}{N} \sum_{k=1}^{N} \sum_{i=1}^{M} \mathbf{1}(a_k = a_i) \log(p(a_i | r_k))
    \label{eq:cce}
\end{equation}

\noindent where $a_k$ is the sequence of originally transmitted 4PAM symbols at TX at time $k$, $N$ is the number of symbols transmitted, $M = 4$ is the modulation order, $r_k$ is the downsampled sample sequence from the DSO, $p(a_i | r_k)$ are the symbol probabilities associated with each of the 4 possible $a_k$ after detection and $\mathbf{1}$ is the indicator function, defined as:

\begin{figure*}[ht!]
    \centering
    \includegraphics[width=\linewidth]{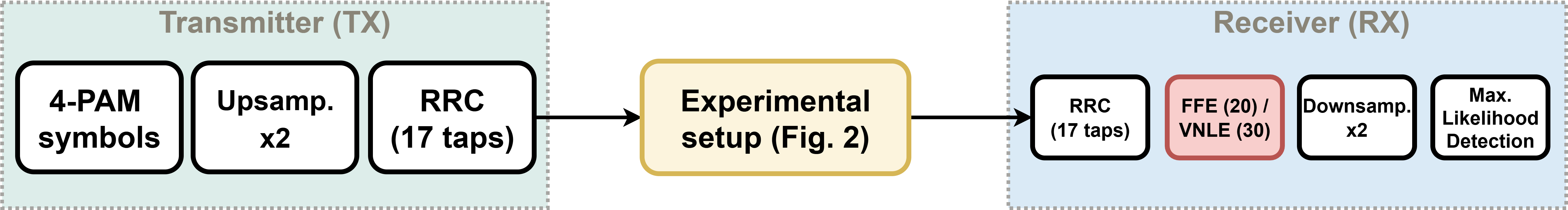}
    \caption{Block diagram of the RX equalization benchmark schemes. The elements optimized by the scheme are highlighted in red. Acronyms: \textbf{4PAM}: 4-Pulse Amplitude Modulation; \textbf{RRC}: root raised cosine; \textbf{FFE}: feedforward equalizer; \textbf{VNLE}: Volterra nonlinear equalizer.}
    \label{fig:dsp_bench}
\end{figure*}

\begin{equation}
    \mathbf{1}(A) = \begin{cases} 1, & \text{if } A \text{ is true} \\ 0, & \text{if } A \text{ is false} \end{cases}
    \label{eq:indicator}
\end{equation}

\noindent where $A$ is an arbitrary logical operation. In practice, $H$ penalizes the uncertainty in the decision of the E2E scheme, i.e. the lower the a-posteriori probability $p(a_i | r_k)$ assigned to the originally transmitted symbol $a_i$ the higher the value of $H$. The main advantage of $H$ as loss function is the inclusion of the decision thresholds in the calculation of the loss while avoiding the use of non-differentiable functions. The minimization of the loss function for the E2E scheme is also performed iteratively based on an Adam optimizer, using a learning rate $\mu_\mathrm{E2E} = 10^{-5}$ and $\beta_1 = 0.999, \beta_2 = 0.99$.

Once the optimization phase is complete, the E2E scheme is validated on the experimental setup to verify its performance. This is done by applying the DSP, $P_\mathrm{RF}$ and $I_\mathrm{bias}$ configurations obtained in the offline optimization to the experimental setup. Given that the optimized ${x_\mathrm{TX}}$ represents voltage levels, the only change applied to the TX DSP on the experimental setup is the downsampling of ${x_\mathrm{TX}}$ from 80 to 65~GSa/s, in order to match the sampling frequency of the AWG. The metric utilized to measure the performance of the E2E scheme is SER. \flag{This is due to the per-symbol nature of the proposed E2E optimization scheme, which allows evaluating signal quality regardless of the bit mapping employed.} The SER calculation on the E2E scheme is performed based on $a_k$ and $r_k$, following:

\begin{equation}
    \mathrm{SER} = \frac{1}{N} \sum_{k=1}^{N} \mathbf{1}(a_k \neq \arg\max_{a_i} p(a_i|r_k))
    \label{eq:ser}
\end{equation}

\subsection{Benchmark RX-side equalization schemes} \label{subsec:bench}

The performance of the E2E scheme is compared to two (FFE, VNLE) RX-side equalization benchmark schemes, giving context on the performance gains obtained by using E2E learning. As in the case of the E2E scheme, the transmission link is based on the experimental setup in \cref{fig:modelsetup}b. The DSP structure of the benchmark schemes is represented in \cref{fig:dsp_bench}. The TX structure of both FFE and VNLE benchmark schemes is identical: both feature an equispaced 4PAM mapper, 2-SpS upsampling, and a 17-tap RRC cosine filter with roll-off factor $\alpha = 0.1$. The obtained discrete sequence is then linearly upsampled to 65~GSa/s for digital-to-analog conversion. Given that $P_\mathrm{RF}$ is not numerically optimized in the case of the benchmark setups, $P_\mathrm{RF}$ is swept in the range $[-4, 2]$~dBm. $I_\mathrm{bias}$ is set to the same value as the one learned by the E2E scheme to allow a fair comparison. On the RX side, the signal is first downsampled to 2 SpS. The two benchmark setups differ only in the RX equalization method. While the linear equalization scheme uses a 20-tap FIR FFE, the nonlinear one is based on a 2$^\mathrm{nd}$ order VNLE, using 20 taps for the 1$^\mathrm{st}$ order kernel and 10 taps for the nonlinear kernel. The 20 linear taps correspond to the total length of the pulse shaping and EQ filters of the E2E setup. \flag{The FFE-based structure of the E2E setup gives it equivalent $\mathcal{O}(N)$ complexity to the FFE benchmark setup.} By employing both linear and nonlinear equalization as benchmarks, we quantify the impact of nonlinearities on the link's performance across $R_s$ and $L$. After equalization, the signal is downsampled to 1 SpS and detected using 1-tap maximum likelihood detection (MLD). Given that the MLD provides hard decision thresholding, the SER is calculated as the fraction of detection errors to the total number of symbols $N$.

\begin{figure*}[hb]
    \centering
    \includegraphics[width=\linewidth]{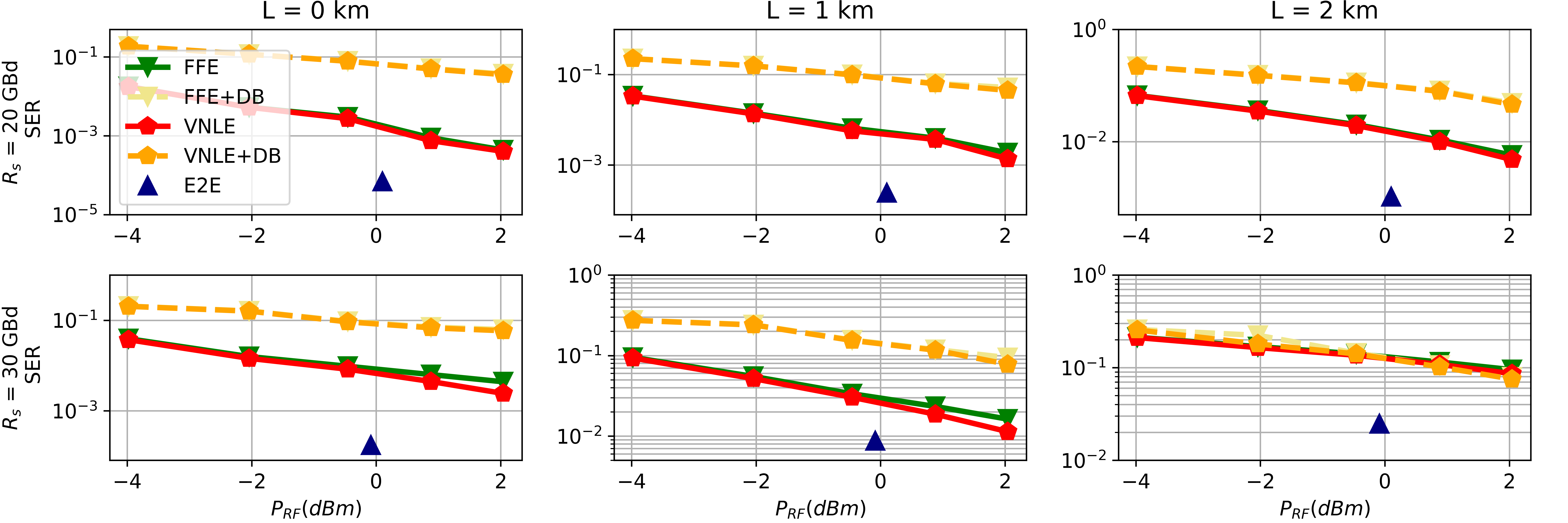}
    \caption{SER performance vs modulation RF power $P_\mathrm{RF}$ of the E2E and benchmark schemes for $R_{s} = 20$~GBd (top) and 30~GBd (bottom) and $L=\{0, \; 1, \; 2\}$~km (left, center, right).}
    \label{fig:serresults}
\end{figure*}

Building on the FFE and VNLE benchmarks, we introduce two additional benchmark schemes. These new benchmarks employ the same equalizer technology but are distinguished by their use of duobinary (DB) signaling. DB signaling is a well-known partial response signaling scheme used in optical communication systems to improve their robustness to CD \cite{618386, 8418771, 10536155}. This is done by reducing the bandwidth of the transmitted signal adding a controlled amount of ISI between neighboring symbols. This also reduces signal attenuation due to hardware bandwidth limitation at the expense of increasing the decoding complexity and the risk of error propagation. DB signaling is an interesting benchmark because its inherent spectral efficiency can mitigate the tight bandwidth constraints of DMLs, offering a low-complexity method to improve performance in cost-sensitive optical links. DB signaling is characterized by the use of an encoding block with impulse response $h_\mathrm{DB} = [1, 1]$ before pulse shaping. This encoding block introduces the desired correlation between symbols leading to ISI while compressing the bandwidth of the signal. At the RX side, the DSP structure is identical to the one used for standard 4PAM with the exception of the MLSE, that needs to account for the correlation introduced between symbols. When paired with 4PAM symbols, the DB scheme is usually referred to as duo-quaternary (DB-4PAM). Considering the larger constellation space due to DB encoding, MLSE is performed considering all the possible symbol combinations (7 in a DB-4PAM scheme). DB-4PAM has an effective constellation size of 7 symbols because each symbol in the output sequence is the sum of the current 4PAM symbol and the previous one, and the combination of these four input levels (for example, $\{-3, -1, 1, 3\}$) results in seven unique output levels ($\{-6, -4, -2, 0, 2, 4, 6\}$). Although inverting the impulse response of the DB encoding is theoretically viable, it is often avoided in practical settings in order to minimize error propagation \cite{10536155}.

\section{Experimental transmission results} \label{sec:results}

The performance comparison of the E2E and RX EQ benchmark schemes includes 2 different baudrates (20 and 30~GBd) and 3 different transmission distances (0, 1, 2~km) for a total of 6 cases. \flag{The baudrates are chosen to approximately match the bandwidth of the laser at low $I_\mathrm{bias}$.} The hyperparameters in the optimization of the E2E scheme (learning rate, number of FIR filter taps at TX and RX) are simultaneously fine-tuned using multi-dimensional grid search. The reported results correspond to the scheme achieving minimum SER when tested on the experimental setup. This process is conducted for every $R_s$ and $L$ independently. When calculated from the experimental setup, the SER is calculated based on a sequence of $10^6$ symbols. The RX EQ benchmarks are trained offline using experimentally obtained sequences. 
The reported SER results are obtained using previously unseen sequences for a total length of $10^6$ symbols. The experimentally obtained SER results for $R_{s} = \{20, \; 30\}$~GBd and $L=\{0, \; 1, \; 2\}$~km with respect to $P_\mathrm{RF}$ are plotted in \cref{fig:serresults}.

The FFE+DB and VNLE+DB labels denote the benchmark schemes using FFE and VNLE equalizers while using DB-4PAM signaling on the TX side. The SER results differ significantly depending on the transmission distance studied. The B2B transmission ($L=0$~km) SER results transmitting at a rate $R_s = 20$~GBd show a clear performance advantage for the E2E scheme with respect to the RX EQ schemes ($6.97 \times 10^{-5}$ vs $4.01 \times 10^{-4}$ within the allowed $P_\mathrm{RF}$). The E2E scheme outperforms its RX EQ counterparts in terms of SER while lowering the required input RF power by almost 2~dB. The minor difference between the FFE and VNLE schemes in \cref{fig:serresults} suggests that the DML is operating in a mostly linear regime, where the second order Volterra terms have a negligible impact. The performance of the DB schemes (FFE+DB and VNLE+DB) in this scenario is worse than the standard signaling counterparts, as expected. This is due to their increased effective constellation size increasing the error probability. On the other hand, CD does not impact performance in a B2B transmission, and therefore the strengths of DB signaling cannot be exploited. 

The $L = 0$~km, $R_s = 30$~GBd case shows a similar trend. The E2E scheme is again the best performing one overcoming the VNLE scheme ($1.74 \times 10^{-4}$ vs $2.40 \times 10^{-3}$ SER). The $P_\mathrm{RF}$ of the E2E scheme is slightly lower than in the $R_s = 20$~GBd case, likely to compensate for the relatively longer rise time of the laser in relation to the symbol rate. However, the VNLE scheme does achieve slightly better performance than the FFE at high $P_\mathrm{RF}$. This is likely due to the higher nonlinear distortion at higher baudrates, that gives the quadratic coefficients of the VNLE an advantage. The DB benchmark schemes achieve similar SERs as in the 20~GBd case.

\begin{figure*}[hb!]
    \centering
    \includegraphics[width=\linewidth]{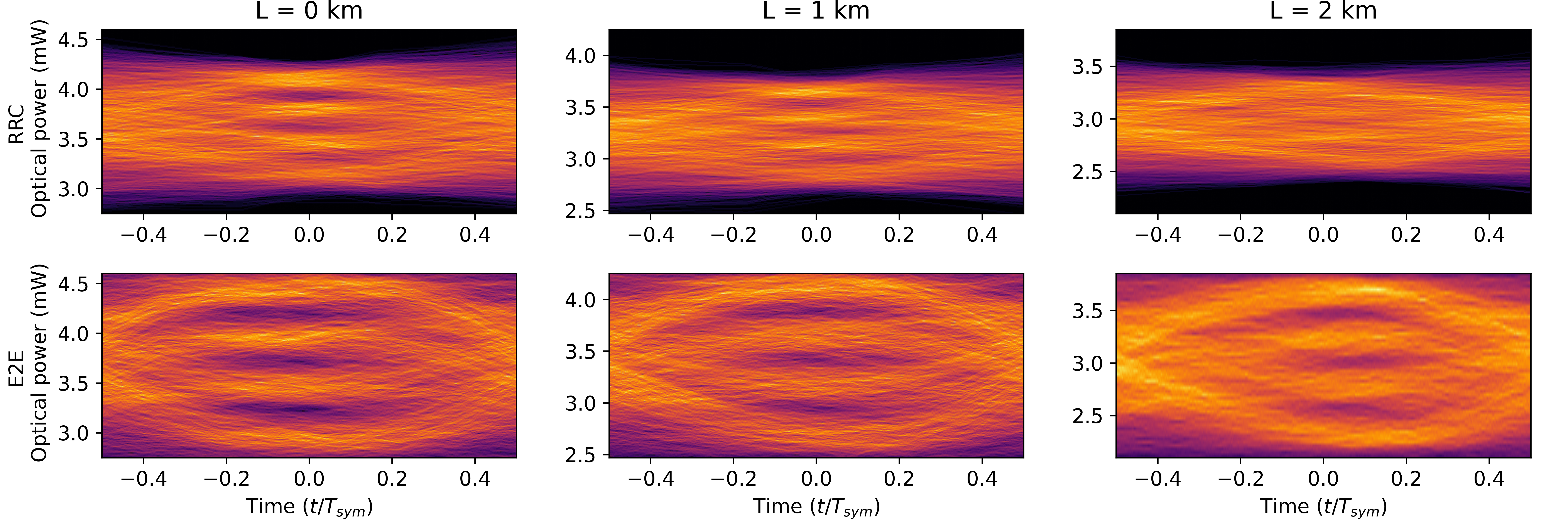}
    \caption{Eye diagrams of the electrical RX-side signals before equalization for the RRC-shaped standard-signaling benchmark schemes (top) and the proposed E2E scheme (bottom) for $L=\{0, \; 1, \; 2\}$~km (left, center, right).}
    \label{fig:eyediags}
\end{figure*}

Looking at the $L=1$~km case, where the CD and chirp interaction starts to have an impact on the performance, the SER performance difference between schemes stay relatively the same. The E2E scheme achieves the lower SER again with respect to the VNLE ($2.44 \times 10^{-4}$ vs $1.36 \times 10^{-3}$ at 20~GBd and $8.75 \times 10^{-3}$ vs $1.13 \times 10^{-2}$ at 30~GBd). In absolute terms, the 20~GBd error rates remain similar to B2B transmission, while the 30~GBd ones are about an order of magnitude higher for all non-DB schemes. This advantage may be due to the fact that low-baudrate signals are more resilient to CD \cite{Zhang:18}. The DB-based schemes show virtually no performance difference with the B2B case at both $R_s$, thanks to their 50\% smaller signal bandwidth. 

When increasing the transmission distance $L$ to 2~km, the difference in performance between the E2E and RX EQ schemes becomes more pronounced at both symbol rates. For $R_s = 20$~GBd, the E2E outperforms the benchmarks schemes, including the VNLE ($1.04 \times 10^{-3}$ vs $4.78 \times 10^{-3}$). Nonetheless, the E2E performs significantly worse than for $L=1$~km ($1.04 \times 10^{-3}$ vs $2.44 \times 10^{-4}$), while the non-DB benchmark schemes suffer an order of magnitude in SER penalty. The DB schemes maintain an approximately equal performance to the standard signaling benchmark schemes, showing their resilience to CD despite the increased transmission distance. For $R_s = 30$~GBd this trend becomes even more pronounced. The E2E scheme remains the best performing one over the VNLE+DB scheme ($2.49 \times 10^{-2}$ vs $7.51 \times 10^{-2}$), but its SER raises above the $10^{-2}$ threshold. In this case, the DB schemes show a slightly superior performance than their non-DB counterparts, although the SER of all the benchmark schemes stay close to $10^{-1}$, which is considered very high. Looking at the $P_\mathrm{RF}$ across the different $L$, it is apparent that the E2E schemes maintain $P_\mathrm{RF}$  constant across the considered values of $L$, even though it changes with $R_s$. This suggests that the laser dynamics have a larger impact on performance than the span length utilized, and therefore the optimized $P_\mathrm{RF}$ is affected mainly by $R_s$.

The performance improvement of the E2E scheme over the RX EQ schemes can be better understood by inspecting the waveforms generated by each scheme. \cref{fig:eyediags} shows the eye diagrams of the electrical signals generated by the E2E and the RRC-shaped signals used in the benchmark FFE and VNLE schemes before RX-side equalization. The figure shows the 20~GBd signals for a transmission distance of 0, 1 and 2~km, from left to right. The 30~GBd signals cannot be plotted due to lack of time resolution in the DSO. 

\begin{figure*}[t]
    \centering
    \includegraphics[width=\linewidth]{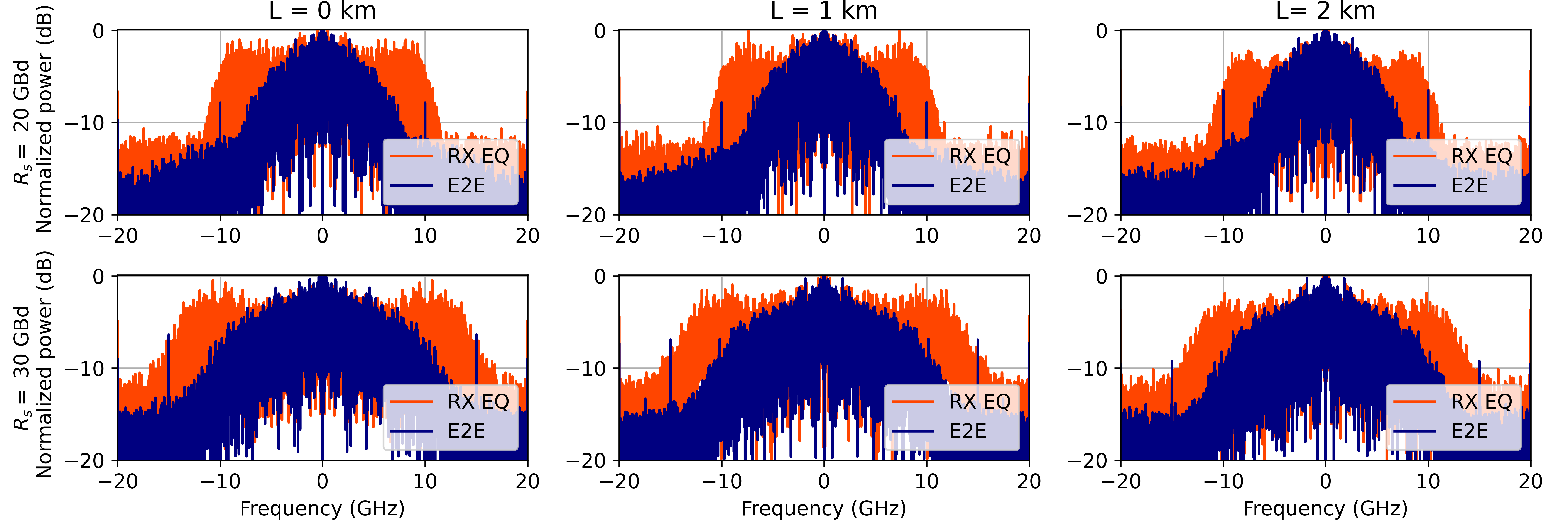}
    \caption{Spectra of the electrical RX-side signals before equalization for the RRC-shaped standard-signaling benchmark schemes and the proposed E2E scheme for $R_{s} = 20$~GBd (top) and 30~GBd (bottom) and $L=\{0, \; 1, \; 2\}$~km (left, center, right).}
    \label{fig:spectra}
\end{figure*}

Looking at the $L = 0$~km case, it becomes apparent that the eye opening of the E2E signal is larger than its counterpart. Firstly, the E2E scheme's signal shows a larger distance between adjacent levels than the RRC-shaped signal. Additionally, the extinction ratio between the highest and lowest power symbols is also higher. Given that both signals are subject to the same power constraints at the TX side, the advantage of the E2E scheme over the RRC-shaped signal must come from the constellation and pulse shaping. This can also be seen in the eye width of the pulses in the E2E signal compared to the RRC-shaped one. While the RRC-shaped signal decays rapidly toward the center of the eye, the E2E signal decays noticeably slower. Another interesting difference is the eye skew in both signals. The symbols in the RRC-shaped signal are clearly not horizontally aligned, despite the practical absence of CD. This is likely due to gain compression in the laser cavity, that leads to some symbols having a longer rise time than others. This effect is mostly absent in the E2E signal, which could also explain the lower error rate in the latter. A similar trend can be seen in the 1 and 2~km transmission cases. The chirp-CD interaction can induce power fading in both signals, reducing their SNR at the RX side. However, they are not impacted in the same way.  The RRC-shaped signals suffers severe eye closure at 1~km and its eye becomes indistinguishable at 2~km. On the other hand, the impact of the higher transmission distance on the E2E signal is significantly lower. While the signal is attenuated as the transmission distance increases, its 4 levels remain distinguishable even for $L=2$~km, explaining the SER results in \cref{fig:serresults}. \flag{The difference in amplitude between the three eye diagrams in \cref{fig:eyediags} (top) also shows the substantial power fading introduced by the interaction between CD and chirp as $L$ is increased, which is less pronounced for the E2E scheme.} The advantage of the E2E scheme points again toward an efficient level tuning and pulse shape design of the E2E signal with respect to the RRC-shaped signal, allowing it to deliver better performance in the presence of CD. 

The impact of learnable pulse shaping in the E2E scheme can be observed in \cref{fig:spectra}, where the spectra of the electrical E2E signal and the RRC-shaped signal at RX are plotted for $R_s = 20$ and 30~GBd and the 3 studied transmission distances (0, 1, 2~km). Throughout all the studied symbol rates and transmission distances, it becomes apparent that the E2E scheme is able to compress its bandwidth with respect to its RRC-shaped counterpart. This compression may explain the lower SER and higher eye opening achieved by the E2E scheme. A lower signal bandwidth allows to adapt to the limited modulation bandwidth of the DML, while it generally makes the signal more resilient to CD. The 10-dB bandwidth compression increases from $L = 0$~km (32\% at 20~GBd and 29\% at 30~GBd) to a maximum at $L = 1$~km (34\% and 30\%), before decreasing at $L = 2$~km (32\% and 24\%). This reduction may be due to the increased ISI introduced due to the pulse shaping-induced bandwidth compression. Nevertheless, the E2E scheme achieves much lower error rates than the DB-4PAM schemes with RX-only equalization. This shows the ability of the E2E scheme to find trade-offs between several parameters at the same time, improving performance. As introduced in \cref{subsec:bench}, there is a trade-off between bandwidth reduction and ISI when designing a signaling scheme at a certain baudrate. The lower SNR at higher $R_s$ may dissuade the E2E scheme from introducing as much ISI as in the $L=0$~km case, leading to a lower compression rate. This suggests that the surrogate model is capturing the impact of the fiber channel on the received signal successfully, given that the E2E scheme is able to adapt to the conditions imposed by the different transmission distances and baudrates.

\begin{figure*}[t]
    \centering
    \includegraphics[width=\linewidth]{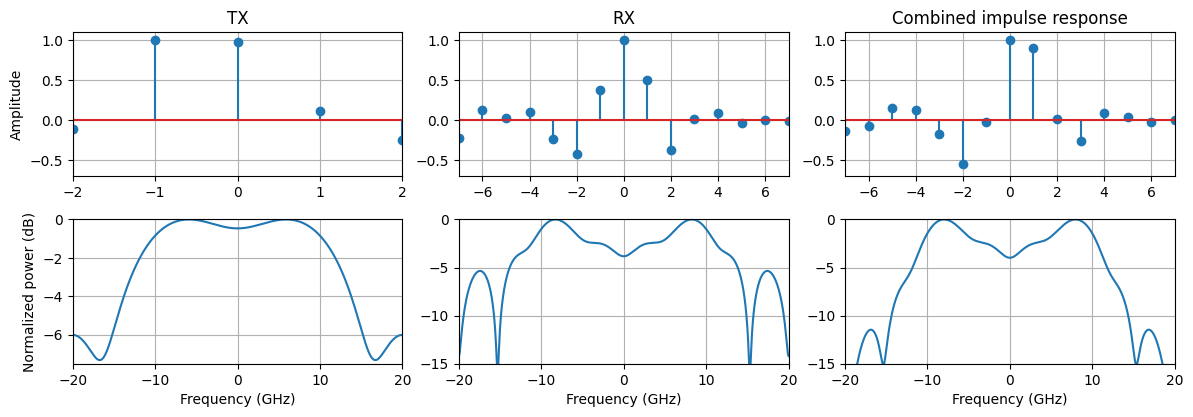}
    \caption{Impulse responses (top) and transfer functions (bottom) of the filters learned by the E2E scheme for $R_s = 20$~GBd and $L = 0$~km. The filters correspond to the TX-side pulse shaper (left), RX-side equalizer (center) and the convolution of pulse shaper and equalizer (right).}
    \label{fig:ae_analysis_0_0}
\end{figure*}

Analyzing the spectra of received signals is a powerful method for evaluating a system's performance. However, the interpretation of these spectra is not always straightforward. This complexity is due to the interplay of all the linear and nonlinear effects contributed by each component in the link. The use of linear FIR filters in the E2E scheme makes it possible to analyze the employed filters at TX and RX using their transfer function, allowing to interpret their individual contribution on the obtained signal. It must be noted that the nonlinear response of laser and PD still plays a significant role in the overall response shown in \cref{fig:spectra}, and therefore the linear time-invariant system analysis cannot be applied to the whole system.

\cref{fig:ae_analysis_0_0} shows the obtained 2-SpS TX (pulse shaping) and RX (equalizing) filters impulse responses optimized by the E2E at $R_s = 20$~GBd and $L = 0$. The combined impulse response of the TX and RX filters, which is the convolution of the two filters, is also shown. The spectra of each impulse response are shown right below them. The 5-tap time-domain TX filter shows resemblance to traditional rectangular pulse shaping, with two identical samples at the center of the filter. However, the rightmost sample has an amplitude of -0.2, introducing ISI between neighboring samples. This rightmost sample may therefore play a role in the performance advantage of the E2E, balancing intentional signal degradation with bandwidth compression. The spectrum of the TX filter, despite its low order, also shows interesting insight. The filter shows attenuation at low frequencies, exactly where the small-signal response of the laser is higher in \cref{fig:l_i_h21}b. This may serve as signal predistortion, increasing the spectral flatness of the signal. The TX filter shows low out-of-band attenuation, mainly due to its low order. The RX filter shows a resemblance to RRC filters. Although the filter impulse response is mostly symmetric, it shows discrepancy at samples $+/-3$. This relates to the compensation of the ISI introduced at the TX filter. The spectrum of the RX filter, as in the case of the TX filter, shows attenuation at lower frequencies. The first spectral notch of the RX filter is at around 15~GHz, coinciding with the 3-dB bandwidth of the laser at the selected $I_\mathrm{bias}$. The combined impulse response of TX and RX filters, similarly to the RX filter alone, shows a raised-cosine like shape, with the zero crossings displaced. It also shows a certain symmetry with the aforementioned exception at positions $+/-3$. The combined impulse response shows a high resemblance to the RX spectrum with higher low-frequency and out-of-band attenuation.

\section{Conclusion} \label{sec:conc}
 We have presented and experimentally validated an end-to-end learning approach for optimizing directly modulated laser-based systems. Our method leverages a differentiable, data-driven surrogate model to enable gradient propagation between the transmitter and receiver. We demonstrated offline optimization of digital signal processing, bias current, and modulation radiofrequency power using the aforementioned surrogate model. Across the symbol rates and transmission distances tested experimentally, our end-to-end scheme achieved superior error performance compared to receiver-side equalization. Furthermore, it reduced modulation radiofrequency power by 2 dB and decreased system bandwidth utilization by over 24\%.

\section{Acknowledgements} The Villum VI-POPCOM (no. VIL54486) and Villum YIP OPTIC-AI (no. VIL29334) projects are acknowledged.

\bibliography{refs}

@article{SIEGELMANN1995132,
title = {{On the Computational Power of Neural Nets}},
journal = {Journal of Computer and System Sciences},
volume = {50},
number = {1},
pages = {132-150},
year = {1995},
issn = {0022-0000},
doi = {https://doi.org/10.1006/jcss.1995.1013},
author = {H.T. Siegelmann and E.D. Sontag}
}

@ARTICLE{10273773,
  author={Ruiz, Marc and Hernandez, Jose Alberto and Quagliotti, Marco and Hugues Salas, Emilio and Riccardi, Emilio and Rafel, Albert and Velasco, Luis and Gonzalez de Dios, Oscar},
  journal={Journal of Optical Communications and Networking}, 
  title={{Network Traffic Analysis under Emerging Beyond-5G Scenarios for Multi-Band Optical Technology Adoption}}, 
  year={2023},
  volume={15},
  number={11},
  pages={F36-F47},
  doi={10.1364/JOCN.492128}}

@Article{app13148125,
AUTHOR = {Xie, Yao and He, Peili and Li, Wei and Li, Na},
TITLE = {{A Decision Feedback Equalization Algorithm Based on Simplified Volterra Structure for PAM4 IM-DD Optical Communication Systems}},
JOURNAL = {Applied Sciences},
VOLUME = {13},
YEAR = {2023},
NUMBER = {14},
ARTICLE-NUMBER = {8125},
ISSN = {2076-3417},
DOI = {10.3390/app13148125}
}

@ARTICLE{10238466,
  author={Che, Di and Chen, Xi},
  journal={Journal of Lightwave Technology}, 
  title={{Modulation Format and Digital Signal Processing for IM-DD Optics at Post-200G Era}}, 
  year={2024},
  volume={42},
  number={2},
  pages={588-605},
  doi={10.1109/JLT.2023.3311716}}

@ARTICLE{10528667,
  author={Wu, Qi and Xu, Zhaopeng and Zhu, Yixiao and Ji, Tonghui and Ji, Honglin and Yang, Yu and Qiao, Gang and Cheng, Chen and Tang, Jianwei and Zhao, Tianfeng and Liang, Junpeng and Liu, Lulu and Wang, Shangcheng and He, Zhixue and Wei, Jinlong and Zhuge, Qunbi and Hu, Weisheng},
  journal={Journal of Lightwave Technology}, 
  title={{High-Speed Dispersion-Unmanaged DML-Based IM-DD Optics at C-band With Advanced Nonlinear Equalization and Noise Whitening}}, 
  year={2024},
  volume={42},
  number={16},
  pages={5591-5598},
  doi={10.1109/JLT.2024.3399209}}

@ARTICLE{10438718,
  author={Diamantopoulos, Nikolaos-Panteleimon and Fujii, Takuro and Yamaoka, Suguru and Nishi, Hidetaka and Takeda, Koji and Segawa, Toru and Matsuo, Shinji},
  journal={Journal of Lightwave Technology}, 
  title={{16-Channel Directly Modulated Membrane III-V Laser Array on SiO2/Si Utilizing Photon-Photon Resonance}}, 
  year={2024},
  volume={42},
  number={11},
  pages={3997-4005},
  doi={10.1109/JLT.2024.3366532}}

@ARTICLE{10080935,
  author={Pang, Xiaodan and Salgals, Toms and Louchet, Hadrien and Che, Di and Gruen, Markus and Matsui, Yasuhiro and Dippon, Thomas and Schatz, Richard and Joharifar, Mahdieh and Krüger, Benjamin and Pittala, Fabio and Fan, Yuchuan and Udalcovs, Aleksejs and Zhang, Lu and Yu, Xianbin and Spolitis, Sandis and Bobrovs, Vjaceslavs and Popov, Sergei and Ozolins, Oskars},
  journal={Journal of Lightwave Technology}, 
  title={{200 Gb/s Optical-Amplifier-Free IM/DD Transmissions Using a Directly Modulated O-Band DFB+R Laser Targeting LR Applications}}, 
  year={2023},
  volume={41},
  number={11},
  pages={3635-3641},
  doi={10.1109/JLT.2023.3261421}}

@ARTICLE{10323176,
  author={Alam, Md Samiul and Maram, Reza and Shahriar, Kh Arif and Ricciardi, Pasquale and Plant, David V.},
  journal={Journal of Lightwave Technology}, 
  title={{Chirped Managed Laser for Multilevel Modulation Formats: A Semi-Analytical Approach for Efficient Filter Design}}, 
  year={2024},
  volume={42},
  number={7},
  pages={2351-2361},
  doi={10.1109/JLT.2023.3334329}}

@ARTICLE{10041217,
  author={Zhu, Xuyuan and Guo, Jing and Li, Huan and Li, Zhenyu and Zhou, Daibing and Zhao, Lingjuan and Wang, Wei and Liang, Song},
  journal={IEEE Photonics Technology Letters}, 
  title={{High Speed Directly Modulated DFB Lasers Having MQW Based Passive Reflectors}}, 
  year={2023},
  volume={35},
  number={6},
  pages={333-336},
  doi={10.1109/LPT.2023.3243638}}

@ARTICLE{9874980,
  author={Guan, Shijian and Zhang, Yunshan and Zheng, Jilin and Su, Jingyou and Sun, Zhenxing and Lu, Linlin and Fang, Tao and Li, Lianyan and Xiao, Rulei and Shi, Yuechun and Chen, Xiangfei},
  journal={Journal of Lightwave Technology}, 
  title={{Modulation Bandwidth Enhancement and Frequency Chirp Suppression in Two-Section DFB Laser}}, 
  year={2022},
  volume={40},
  number={22},
  pages={7383-7389},
  doi={10.1109/JLT.2022.3203723}}

@ARTICLE{8570831,
  author={Wang, Weiyu and Li, Huanlu and Zhao, Pengchao and Zhang, Zhike and Zang, Dajun and Wang, Cuicui and Li, Liang and Ma, Lin and Liu, Yu and Zhu, Ninghua and Lu, Yuchun},
  journal={Journal of Lightwave Technology}, 
  title={{Advanced Digital Signal Processing for Reach Extension and Performance Enhancement of 112 Gbps and Beyond Direct Detected DML-Based Transmission}}, 
  year={2019},
  volume={37},
  number={1},
  pages={163-169},
  doi={10.1109/JLT.2018.2885707}}

@ARTICLE{10925524,
  author={Yuan, Alan Yi-Lun and Savory, Seb J.},
  journal={Journal of Lightwave Technology}, 
  title={{Data-Driven Modeling of a Direct Detection System With an Electro-Absorption Modulated Laser}}, 
  year={2025},
  volume={43},
  number={12},
  pages={5545-5557},
  doi={10.1109/JLT.2025.3549297}}

@article{Zhang:25,
author = {Qifan Zhang and Shi Jia and Tianhao Zhang and Jinlong Yu},
journal = {Opt. Express},
number = {2},
pages = {2360--2375},
title = {{Accurate Deep Learning based Method for Real-Time Directly Modulated Laser Modeling}},
volume = {33},
month = {Jan},
year = {2025},
doi = {10.1364/OE.549604},
}

@inproceedings{Sudo:21,
author = {Tsurugi Sudo and Yasuhiro Matsui and Glen Carey and Ashish Verma and Ding Wang and Viral Lowalekar and Martin Kwakernaak and Ferdous Khan and Nicholas Dalida and Ronak Patel and Alexander Nickel and Bruce Young and Jimmy Zeng and Yuk Lung Ha and Charles Roxlo},
booktitle = {Optical Fiber Communication Conference (OFC) 2021},
journal = {Optical Fiber Communication Conference (OFC) 2021},
pages = {Tu1B.3},
title = {{Challenges and Opportunities of Directly Modulated Lasers in Future Data Center and 5G Networks}},
year = {2021},
doi = {10.1364/OFC.2021.Tu1B.3},
}

@INPROCEEDINGS{575810,
  author={Bennett, S. and Snowden, C.M. and Iezekiel, S.},
  booktitle={Proceedings of EDMO '96}, 
  title={{Nonlinear Dynamics in Directly Modulated Multiple Quantum Well Laser Diodes}}, 
  year={1996},
  volume={},
  number={},
  pages={102-107},
  doi={10.1109/EDMO.1996.575810}}

@ARTICLE{8259239,
  author={Zhong, Kangping and Zhou, Xian and Huo, Jiahao and Yu, Changyuan and Lu, Chao and Lau, Alan Pak Tao},
  journal={Journal of Lightwave Technology}, 
  title={{Digital Signal Processing for Short-Reach Optical Communications: A Review of Current Technologies and Future Trends}}, 
  year={2018},
  volume={36},
  number={2},
  pages={377-400},
  doi={10.1109/JLT.2018.2793881}}

@inproceedings{Kottke:17,
author = {Christoph Kottke and Christoph Caspar and Volker Jungnickel and Ronald Freund and Mikel Agustin and Nikolay N. Ledentsov},
booktitle = {Optical Fiber Communication Conference},
journal = {Optical Fiber Communication Conference},
pages = {W4I.7},
title = {{High Speed 160 Gb/s DMT VCSEL Transmission Using Pre-equalization}},
year = {2017},
doi = {10.1364/OFC.2017.W4I.7},
}

@ARTICLE{10526441,
  author={Nagarajan, Radhakrishnan and Martino, Agustin and Morero, Damian A. and Patra, Lenin and Lutkemeyer, Christian and Castrillón, Mario A.},
  journal={Journal of Lightwave Technology}, 
  title={{Recent Advances in Low-Power Digital Signal Processing Technologies for Data Center Applications}}, 
  year={2024},
  volume={42},
  number={12},
  pages={4222-4232},
  doi={10.1109/JLT.2024.3399032}}

@ARTICLE{10530894,
  author={Wang, Danshi and Song, Yuchen and Zhang, Yao and Jiang, Xiaotian and Dong, Jiawei and Khan, Faisal Nadeem and Sasai, Takeo and Huang, Shanguo and Lau, Alan Pak Tao and Tornatore, Massimo and Zhang, Min},
  journal={Journal of Lightwave Technology}, 
  title={{Digital Twin of Optical Networks: A Review of Recent Advances and Future Trends}}, 
  year={2024},
  volume={42},
  number={12},
  pages={4233-4259},
  doi={10.1109/JLT.2024.3401419}}

@article{Agrell_2024,
doi = {10.1088/2040-8986/ad261f},
year = {2024},
month = {jul},
volume = {26},
number = {9},
pages = {093001},
author = {Agrell, Erik and Karlsson, Magnus and Poletti, Francesco and Namiki, Shu and Chen, Xi (Vivian) and Rusch, Leslie A and Puttnam, Benjamin and Bayvel, Polina and Schmalen, Laurent and Tao, Zhenning and Kschischang, Frank R and Alvarado, Alex and Mukherjee, Biswanath and Casellas, Ramon and Zhou, Xiang and van Veen, Dora and Mohs, Georg and Wong, Elaine and Mecozzi, Antonio and Alouini, Mohamed-Slim and Diamanti, Eleni and Uysal, Murat},
title = {{Roadmap on Optical Communications}},
journal = {Journal of Optics},
}

@ARTICLE{10124361,
  author={Jovanovic, Ognjen and {Da Ros}, Francesco and Zibar, Darko and Yankov, Metodi P.},
  journal={Journal of Lightwave Technology}, 
  title={{Geometric Constellation Shaping for Fiber-Optic Channels via End-to-End Learning}}, 
  year={2023},
  volume={41},
  number={12},
  pages={3726-3736},
  doi={10.1109/JLT.2023.3276300}}

@ARTICLE{10093964,
  author={Rode, Andrej and Geiger, Benedikt and Chimmalgi, Shrinivas and Schmalen, Laurent},
  journal={Journal of Lightwave Technology}, 
  title={{End-to-End Optimization of Constellation Shaping for Wiener Phase Noise Channels With a Differentiable Blind Phase Search}}, 
  year={2023},
  volume={41},
  number={12},
  pages={3849-3859},
  doi={10.1109/JLT.2023.3265308}}

@ARTICLE{8054694,
  author={O’Shea, Timothy and Hoydis, Jakob},
  journal={IEEE Transactions on Cognitive Communications and Networking}, 
  title={{An Introduction to Deep Learning for the Physical Layer}}, 
  year={2017},
  volume={3},
  number={4},
  pages={563-575},
  doi={10.1109/TCCN.2017.2758370}}

@ARTICLE{10824218,
  author={Miguel Mateos-Ramos, José and Häger, Christian and Furkan Keskin, Musa and Le Magoarou, Luc and Wymeersch, Henk},
  journal={IEEE Transactions on Wireless Communications}, 
  title={{Model-Based End-to-End Learning for Multi-Target Integrated Sensing and Communication Under Hardware Impairments}}, 
  year={2025},
  volume={24},
  number={3},
  pages={2574-2589},
  doi={10.1109/TWC.2024.3522667}}

@inproceedings{Karanov:20,
author = {Boris Karanov and Mathieu Chagnon and Vahid Aref and Domani\c{c} Lavery and Polina Bayvel and Laurent Schmalen},
booktitle = {Optical Fiber Communication Conference (OFC) 2020},
journal = {Optical Fiber Communication Conference (OFC) 2020},
pages = {Th2A.48},
title = {{Concept and Experimental Demonstration of Optical IM/DD End-to-End System Optimization using a Generative Model}},
year = {2020},
doi = {10.1364/OFC.2020.Th2A.48},
}

@inproceedings{Xu:20,
author = {Zhaopeng Xu and Chuanbowen Sun and Tonghui Ji and Honglin Ji and William Shieh},
booktitle = {Optical Fiber Communication Conference (OFC) 2020},
journal = {Optical Fiber Communication Conference (OFC) 2020},
pages = {W2A.45},
title = {{Cascade Recurrent Neural Network Enabled 100-Gb/s PAM4 Short-Reach Optical Link Based on DML}},
year = {2020},
doi = {10.1364/OFC.2020.W2A.45},
}

@ARTICLE{7972948,
  author={Zhu, Ning Hua and Shi, Zhan and Zhang, Zhi Ke and Zhang, Yi Ming and Zou, Can Wen and Zhao, Ze Ping and Liu, Yu and Li, Wei and Li, Ming},
  journal={IEEE Journal of Selected Topics in Quantum Electronics}, 
  title={{Directly Modulated Semiconductor Lasers}}, 
  year={2018},
  volume={24},
  number={1},
  pages={1-19},
  doi={10.1109/JSTQE.2017.2720959}}

@ARTICLE{10059125,
  author={Srinivasan, Muralikrishnan and Song, Jinxiang and Grabowski, Alexander and Szczerba, Krzysztof and Iversen, Holger K. and Schmidt, Mikkel N. and Zibar, Darko and Schröder, Jochen and Larsson, Anders and Häger, Christian and Wymeersch, Henk},
  journal={Journal of Lightwave Technology}, 
  title={{End-to-End Learning for VCSEL-Based Optical Interconnects: State-of-the-Art, Challenges, and Opportunities}}, 
  year={2023},
  volume={41},
  number={11},
  pages={3261-3277},
  doi={10.1109/JLT.2023.3251660}}

@ARTICLE{10261204,
  author={Minelli, Leonardo and Forghieri, Fabrizio and Shao, Tong and Shahpari, Ali and Gaudino, Roberto},
  journal={Journal of Lightwave Technology}, 
  title={{TDECQ-Based Optimization of Nonlinear Digital Pre-Distorters for VCSEL-MMF Optical Links Using End-to-End Learning}}, 
  year={2024},
  volume={42},
  number={2},
  pages={621-635},
  doi={10.1109/JLT.2023.3318295}}

@ARTICLE{10382548,
  author={Hernandez, Sergio and Jovanovic, Ognjen and Peucheret, Christophe and Da Ros, Francesco and Zibar, Darko},
  journal={IEEE Photonics Technology Letters}, 
  title={{Differentiable Machine Learning-Based Modeling for Directly-Modulated Lasers}}, 
  year={2024},
  volume={36},
  number={4},
  pages={266-269},
  doi={10.1109/LPT.2024.3350993}}

@article{Hernandez:24,
author = {Hernandez, Sergio and Peucheret, Christophe and {Da Ros}, Francesco and Zibar, Darko},
journal = {Journal of Optical Communications and Networking},
number = {8},
pages = {D29--D43},
publisher = {Optica Publishing Group},
title = {{End-to-End Optimization of Optical Communication Systems based on Directly Modulated Lasers}},
volume = {16},
month = {Aug},
year = {2024},
doi = {10.1364/JOCN.522761},
}

@ARTICLE{9513589,
  author={Jovanovic, Ognjen and Yankov, Metodi P. and Da Ros, Francesco and Zibar, Darko},
  journal={Journal of Lightwave Technology}, 
  title={{Gradient-Free Training of Autoencoders for Non-Differentiable Communication Channels}}, 
  year={2021},
  volume={39},
  number={20},
  pages={6381-6391},
  doi={10.1109/JLT.2021.3103339}}

@inproceedings{Hernandez:25,
author = {Hernandez, Sergio and Peucheret, Christophe and {Da Ros}, Francesco and Zibar, Darko},
booktitle = {Optical Fiber Communication Conference (OFC) 2025},
journal = {Optical Fiber Communication Conference (OFC) 2025},
pages = {W4H.3},
title = {{Experimental Demonstration of End-to-End Optimization for Directly Modulated Laser-based IM/DD Systems}},
year = {2025},
doi = {10.1364/OFC.2025.W4H.3},
}

@ARTICLE{618386,
  author={Yonenaga, K. and Kuwano, S.},
  journal={Journal of Lightwave Technology}, 
  title={{Dispersion-Tolerant Optical Transmission System using Duobinary Transmitter and Binary Receiver}}, 
  year={1997},
  volume={15},
  number={8},
  pages={1530-1537},
  doi={10.1109/50.618386}}

@ARTICLE{10536155,
  author={Hu, Qian and Borkowski, Robert},
  journal={Journal of Lightwave Technology}, 
  title={{High-Speed IM/DD Transmission Using Partial Response Signaling With Precoding and Memoryless Decoding}}, 
  year={2024},
  volume={42},
  number={11},
  pages={4038-4047},
  doi={10.1109/JLT.2024.3402688}}

@ARTICLE{8418771,
  author={Zhang, Hong-Bo and Jiang, Ning and Zheng, Zhi and Wang, Wen-Qin},
  journal={IEEE Photonics Journal}, 
  title={{Experimental Demonstration of FTN-NRZ, PAM-4, and Duobinary Based on 10-Gbps Optics in 100G-EPON}}, 
  year={2018},
  volume={10},
  number={5},
  pages={1-13},
  doi={10.1109/JPHOT.2018.2858804}}

@ARTICLE{10839011,
  author={Nielsen, Søren Føns and {Da Ros}, Francesco and Schmidt, Mikkel N. and Zibar, Darko},
  journal={Journal of Lightwave Technology}, 
  title={{End-to-End Learning of Transmitter and Receiver Filters in Bandwidth Limited Fiber Optic Communication Systems}}, 
  year={2025},
  volume={43},
  number={8},
  pages={3749-3760},
  doi={10.1109/JLT.2025.3528542}}

@article{Zhang:18,
author = {Kuo Zhang and Qunbi Zhuge and Haiyun Xin and Weisheng Hu and David V. Plant},
journal = {Optics Express},
number = {26},
pages = {34288-34304},
title = {{Performance comparison of DML, EML and MZM in Dispersion-Unmanaged Short Reach Transmissions with Digital Signal Processing}},
volume = {26},
month = {Dec},
year = {2018},
doi = {10.1364/OE.26.034288},
}
\bibliographystyle{IEEEtran}

\vfill

\end{document}